\documentclass[prl,twocolumn,amsmath,amssymb]{revtex4-2}
\usepackage{comment}
\usepackage{color}
\usepackage{graphicx}
\usepackage{subfigure}
\usepackage{dcolumn}
\usepackage{bm}
\usepackage{amsmath}
\usepackage{braket}
\usepackage{diagbox, eqparbox, hhline}
\usepackage[detect-none]{siunitx}
\usepackage[detect-none]{siunitx}
\usepackage{graphicx}
\usepackage{pgfplots}
\usepgfplotslibrary{fillbetween}
\usepackage{adjustbox}
\usepackage{upgreek}

\pgfplotsset{compat=1.16} 
\begin{document}

\title{Superconductivity and topological behavior in gallenene}

\author{Mikhail Petrov}
\affiliation{Department of Physics and NANOlab Center of Excellence, University of Antwerp, Groenenborgerlaan 171, B-2020 Antwerp, Belgium}

\author{Jonas Bekaert}
\email{jonas.bekaert@uantwerpen.be}
\affiliation{Department of Physics and NANOlab Center of Excellence, University of Antwerp, Groenenborgerlaan 171, B-2020 Antwerp, Belgium}

\author{Milorad V. Milo\v{s}evi\'c}
\email{milorad.milosevic@uantwerpen.be}
\affiliation{Department of Physics and NANOlab Center of Excellence, University of Antwerp, Groenenborgerlaan 171, B-2020 Antwerp, Belgium}

\begin{abstract}
Among the large variety of two-dimensional (2D) materials discovered to date, elemental monolayers that host superconductivity are very rare. Using \textit{ab initio} calculations we show that recently synthesized gallium monolayers, coined gallenene, are intrinsically superconducting through electron-phonon coupling. We reveal that Ga-100 gallenene, a planar monolayer isostructural with graphene, is the structurally simplest 2D superconductor to date, furthermore hosting topological edge states due to its honeycomb structure. Our anisotropic Eliashberg calculations show distinctly three-gap superconductivity in Ga-100, in contrast to the alternative buckled Ga-010 gallenene which presents a single anisotropic superconducting gap. Strikingly, the critical temperature ($T_c$) of gallenene is in the range of $7-10$ K, exceeding the $T_c$ of bulk gallium from which it is exfoliated. Finally we explore chemical functionalization of gallenene with hydrogen, and report induced multigap superconductivity with an enhanced $T_c$ in the resulting gallenane compound.
\end{abstract}

\date{\today}
\pacs{Valid PACS appear here}

\maketitle

\paragraph{Introduction} In the postgraphene era,  a major focus of research on 2D systems are the elemental main-group materials. Being widely used in their bulk form in modern electronics (especially Si and Ge), the exciting physical properties of their 2D structures are currently being explored \cite{elemental_2d_mat}. Beyond graphene, other elements of group 14 of the periodic system have been shown to form slightly buckled honeycomb structures, yielding topologically insulating behavior in the heavier-element monolayers, such as stanene \cite{Deng2018} and plumbene \cite{PhysRevB.95.125113}. In other groups, buckled hexagonal structures are also common, with a band gap increasing with buckling, e.g., in phosphorene \cite{Liu2014}. A wealth of different polymorphs has been found for borophene, all with extended unit cells -- both planar and non-planar -- and moreover highly dependent on the substrate of choice \cite{https://doi.org/10.1002/anie.201505425,Mannix1513,Feng2016}. In contrast with the majority of elemental monolayers, most borophene polymorphs are metallic, and some could even host superconductivity, as recently predicted by \textit{ab initio} calculations \cite{PhysRevB.98.134514}.

\begin{figure*}[t] 
\centering 
\includegraphics[width=\textwidth]{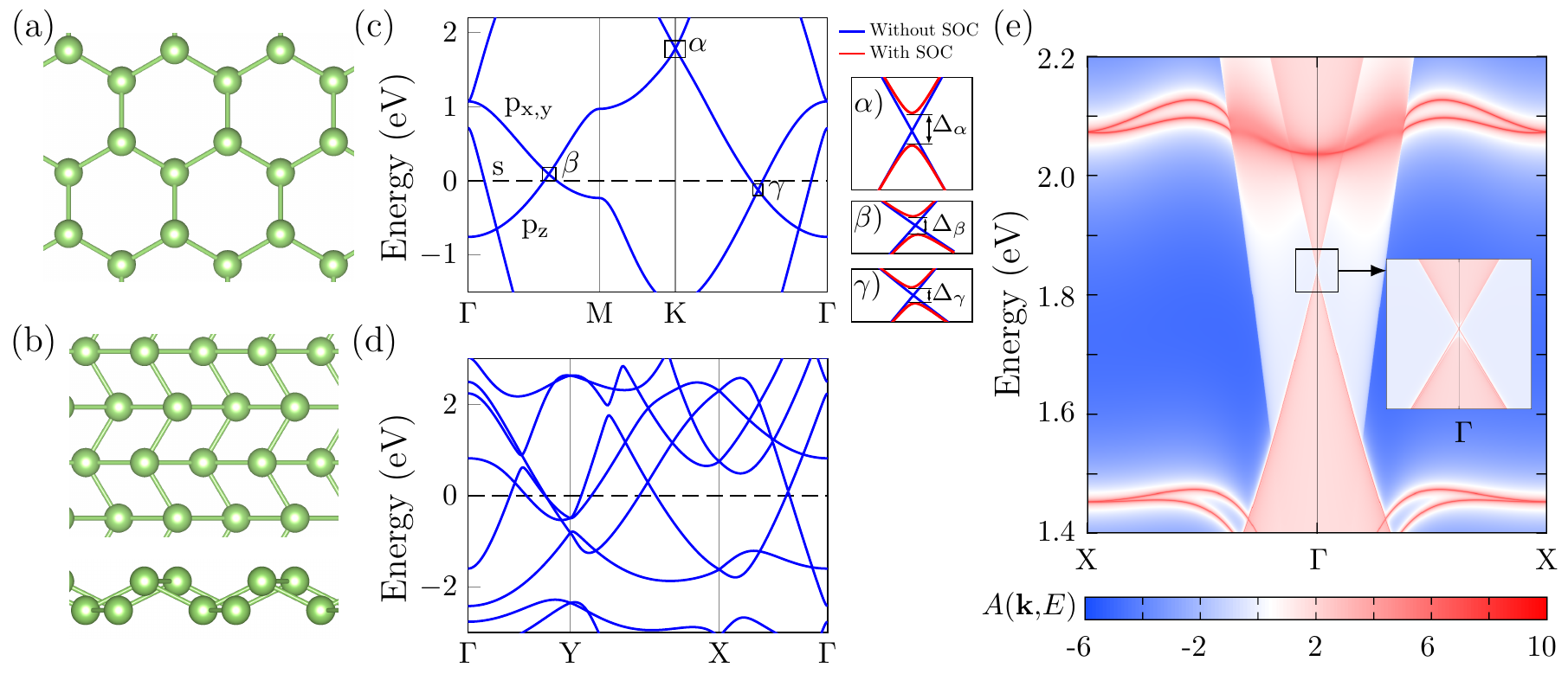}
\caption{Crystal structures of (a) Ga-100 and (b) Ga-010. (c) Electronic structure of 5$\%$ strained Ga-100 (with the Fermi level at 0), showing three Dirac cones indicated by $\alpha$, $\beta$ and $\gamma$. The effect of spin-orbit coupling (SOC) on the cones is depicted in the zoom-ins on the right of the panel, showing the opening of gaps due to SOC ($\Delta_\alpha$, $\Delta_\beta$ and $\Delta_\gamma$). (d) Electronic structure of  2$\%$ strained Ga-010. (e) Surface spectral function of an armchair Ga-100 ribbon. The inset details the $\alpha$ cone (folded to $\Gamma$ in the Brillouin zone of the ribbon). The highest values of $A(\textbf{k},E)$ correspond to the most edge-like states.} 
\label{fig1} 
\end{figure*}

Gallenene, a monolayer of elemental gallium, is the latest addition to the family of elemental 2D materials \cite{Kochate1701373}. Atomically-thin samples of gallenene can be obtained from the bulk $\alpha$-phase of gallium using the solid-melt exfoliation technique, taking advantage of the exceptionally low melting temperature of gallium. This novel method allows to transfer the gallenene structures onto diverse substrates (e.g., GaN, GaAs, Si, and Ni) \cite{Kochate1701373}. Two distinct gallenene phases have thus been produced: Ga-100, named after the corresponding bulk plane, which adopts a planar honeycomb structure (see Fig.~\ref{fig1}(a)), and Ga-010, a two-atom thick ridged structure (see Fig.~\ref{fig1}(b)). Both gallenene phases are metallic, and robustly so with respect to strain and the interaction with different substrates \cite{Kochate1701373,C8CP05280H}. 

Many different gallium allotropes are known to harbor superconductivity. Both the bulk $\alpha$-phase and the metastable $\beta$-phase are superconducting, with critical temperatures ($T_c$) of 1.1 and 6.0 K respectively \cite{PhysRev.150.315,PhysRevB.97.184517}. Ultrathin Ga films grown epitaxially on GaN have also been found to be superconducting, with a $T_c$ of $4-5$ K \cite{PhysRevLett.114.107003,Xing542}, and are notably the first system where a quantum Griffiths singularity, due to disorder, was observed at the superconductor-metal transition \cite{Xing542}. 

Owing to their robust metallic properties, both gallenene structures are prime candidates to host 2D superconductivity. Compared with established elemental monolayer superconductors which are grown epitaxially, e.g. In and Pb \cite{Zhang2010}, gallenene has the advantage of being in principle transferable onto any substrate, so deviations from its intrinsic structural and electronic properties due to hybridization can be considered minimal. Also, the Ga-100 phase presents a unique case of a planar honeycomb material with metallic -- and thus potentially superconducting -- properties, prone to harbor topological edge states due to the combination of a honeycomb lattice with the sizeable nuclear mass of gallium. Altogether this motivates the core objective of this Letter, to explore the superconducting and topological properties of gallenene.  

\paragraph{Structural, vibrational and electronic properties of gallenene} We first investigated the structural and vibrational properties of both gallenene types in freestanding form using density functional (perturbation) theory (DF(P)T), as implemented in Quantum Espresso \cite{QE-2009,doi:10.1063/5.0005082} (computational details are provided in the Supplemental Material). As in earlier work on gallenene \cite{Kochate1701373}, we found minor instabilities in the phonon band structures of the relaxed structures, that are fully cured by a small amount of biaxial tensile strain -- 5\% and 2\% for Ga-100 and Ga-010 respectively. In what follows we therefore focus on these slightly strained structures. High-resolution electron microscopy on both gallenene types did not indicate significant strain \cite{Kochate1701373}, hence it is likely that substrates fully stabilize the structures without the need to apply additional strain.

Due to their distinct lattice structures the two gallenene phases have very different electronic properties. The electronic bands of Ga-100 around the Fermi level have main contributions stemming from in-plane $p_{x,y}$ states and out-of-plane $p_{z}$ atomic orbitals, as well as a smaller hole pocket formed by $s$ states, as indicated in Fig.~\ref{fig1}(c). This yields a Fermi surface composed of three distinct sheets, depicted in Fig.~\ref{fig3}(a). On the other hand, in Ga-010 the in-plane and out-of-plane orbitals are intermixed, giving rise to a tangled Fermi surface (see Figs.~\ref{fig1}(d) and \ref{fig3}(b)). Ga-100 also has a significantly higher density of states (DOS) at the Fermi level than Ga-010 (0.91 and 0.43 states/eV per atom respectively).

\paragraph{Topological properties of Ga-100 gallenene} In the electronic structure of Ga-100 several Dirac-type band crossings are present, indicated by $\alpha$, $\beta$ and $\gamma$ in Fig.~\ref{fig1}(c), with a potential for topological behavior. To investigate this, we have first considered the effect of spin-orbit coupling (SOC) on the band structure. As shown in Fig.~\ref{fig1}(c), the inclusion of SOC leads to gaps opening in all three Dirac cones. Cone $\alpha$, centered around 1.79 eV above the Fermi level ($E_F$) at the K-point of the Brillouin zone, is Ga-100's direct counterpart of graphene's Dirac dispersion \cite{Kochate1701373}. However, in gallenene it is situated above $E_F$ as Ga has one electron less in its outer shell compared with C. The gap within cone $\alpha$ is $\Delta_\alpha=2.9$ meV, two orders of magnitude larger than the corresponding gap in graphene \cite{PhysRevB.80.235431}, owing to the higher atomic mass of Ga. Crossings $\beta$ and $\gamma$ are located 60 meV above and 130 meV below $E_F$ respectively, and both SOC gaps ($\Delta_\beta$ and $\Delta_\gamma$) are 40 meV, about an order of magnitude larger than $\Delta_\alpha$. In order to determine if any of these gaps can host topological edge states, we calculated the surface spectral function (SSF) \cite{PhysRevB.28.4397} of zigzag and armchair Ga-100 ribbons, using a tight-binding model constructed with Wannier functions \cite{MOSTOFI20142309}, as implemented in WannierTools \cite{WU2018405}. The SSF of the Ga-100 ribbon with armchair edges, depicted in Fig.~\ref{fig1}(e), shows that the bulk $\alpha$ gap closes due to edge states with a linear dispersion (see the inset). Hence, Ga-100 harbors SOC-induced topologically insulating behavior among its unoccupied bands. As the $\alpha$ gap amounts to $\Delta_\alpha=2.9$ meV, the topological behavior will manifest itself below a temperature of $\Delta_\alpha/k_{\mathrm{B}}=34$ K. This is markedly higher than the threshold temperature of 0.28 K predicted for graphene \cite{PhysRevB.80.235431}, as a result of the stronger SOC in gallenene. To further corroborate the topological behavior in Ga-100 we have calculated the $Z_2$ topological invariant \cite{PhysRevB.76.045302} in the vicinity of the  $\alpha$ gap, based on Wannier charge centers \cite{PhysRevB.83.035108,WU2018405}. We found the resulting $Z_2$ number to be 1, which supports our conclusion on the non-trivial topological character of the $\alpha$ cone in Ga-100. An analogous analysis yielded no distinct topological edge states related to the $\beta$ and $\gamma$ gaps.  

\begin{figure}[t]
\centering 
\includegraphics[width=\linewidth]{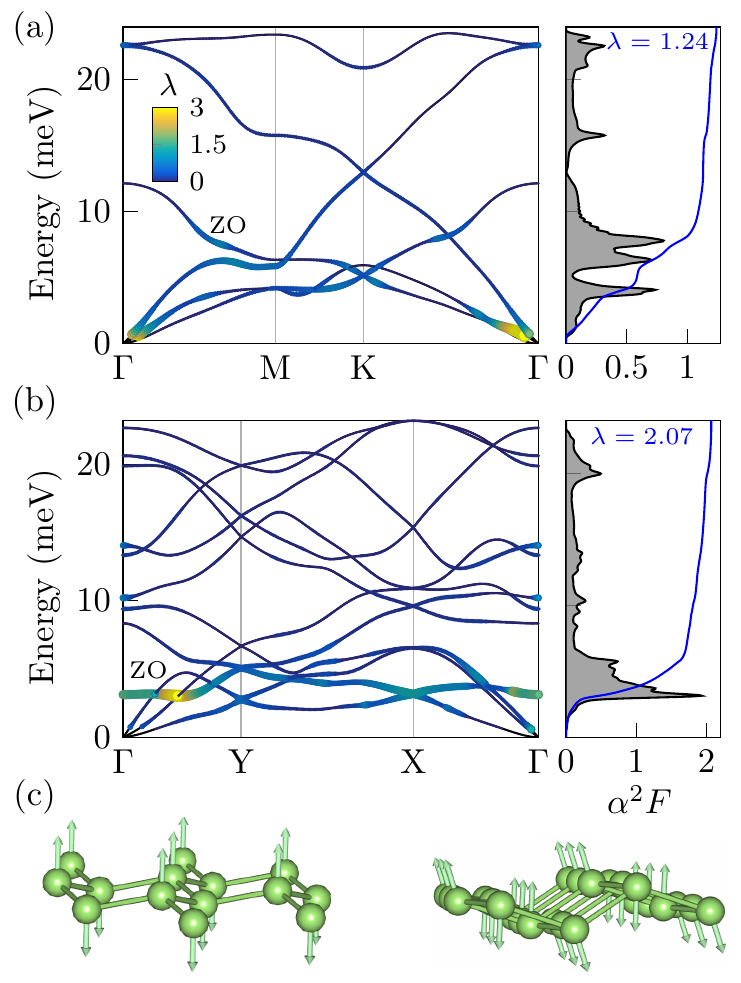}
\caption{Phonon dispersion with the mode-resolved electron-phonon coupling (EPC) indicated by colors, isotropic Eliashberg function $\alpha^2F$, and EPC function $\lambda$ of (a) Ga-100 and (b) Ga-010. The color legend in (a) also applies to (b). (c) ZO phonon mode of Ga-100 (left) and Ga-010 (right), which hosts strong EPC in both structures.}
\label{fig2} 
\end{figure} 

\paragraph{Superconducting properties} As the next step, we calculated the electron-phonon coupling (EPC) in both gallenene structures using the Electron-Phonon Wannier code (EPW) \cite{RevModPhys.89.015003,PONCE2016116}. Our results for the phonon mode-resolved EPC, shown in Fig.~\ref{fig2}(a) and (b), reveal particularly strong coupling in the lower-energy phonon modes in both structures. In particular, the majority of the EPC stems from the optical mode where the Ga atoms oscillate in the out-of-plane direction (i.e., the ZO mode), as depicted in Fig.~\ref{fig2}(c), and within the acoustic modes. The EPC shows significant anisotropy within the Brillouin zone for both monolayers, rather than being strongly peaked at $\Gamma$, as is the case in e.g. MgB$_2$ \cite{PhysRevB.96.094510}. As a result, there is sizeable coupling between electronic states on different parts of the Fermi surface, especially in Ga-010 where the effect is more pronounced and promotes hybridization of the superconducting properties of different bands. In the case of Ga-100 it is interesting to note that the two in-plane optical modes (both belonging to E$_{2\mathrm{g}}$) do not exhibit strong EPC, contrary to what is found for closely related structures, such as doped graphane \cite{PhysRevLett.105.037002} and monolayer MgB$_2$ (which contains a B honeycomb layer) \cite{PhysRevB.96.094510}. This leads to three main peaks in the Eliashberg function $\alpha^2F$ of Ga-100 (Fig.~\ref{fig2}(a)), and one main peak in Ga-010 (Fig.~\ref{fig2}(b)), directly stemming from the ZO mode. The resulting isotropic EPC constant $\lambda$, obtained from the Eliashberg function \cite{ALLEN19831}, amounts to 1.24 for Ga-100 and 2.07 for Ga-010. The ZO mode is lower in energy in the latter (see Fig.~\ref{fig2}(b)), hence this mode contributes more to the total EPC, leading to the larger $\lambda$ value in Ga-010.

\begin{figure}[b] 
\centering 
\includegraphics[width=\linewidth]{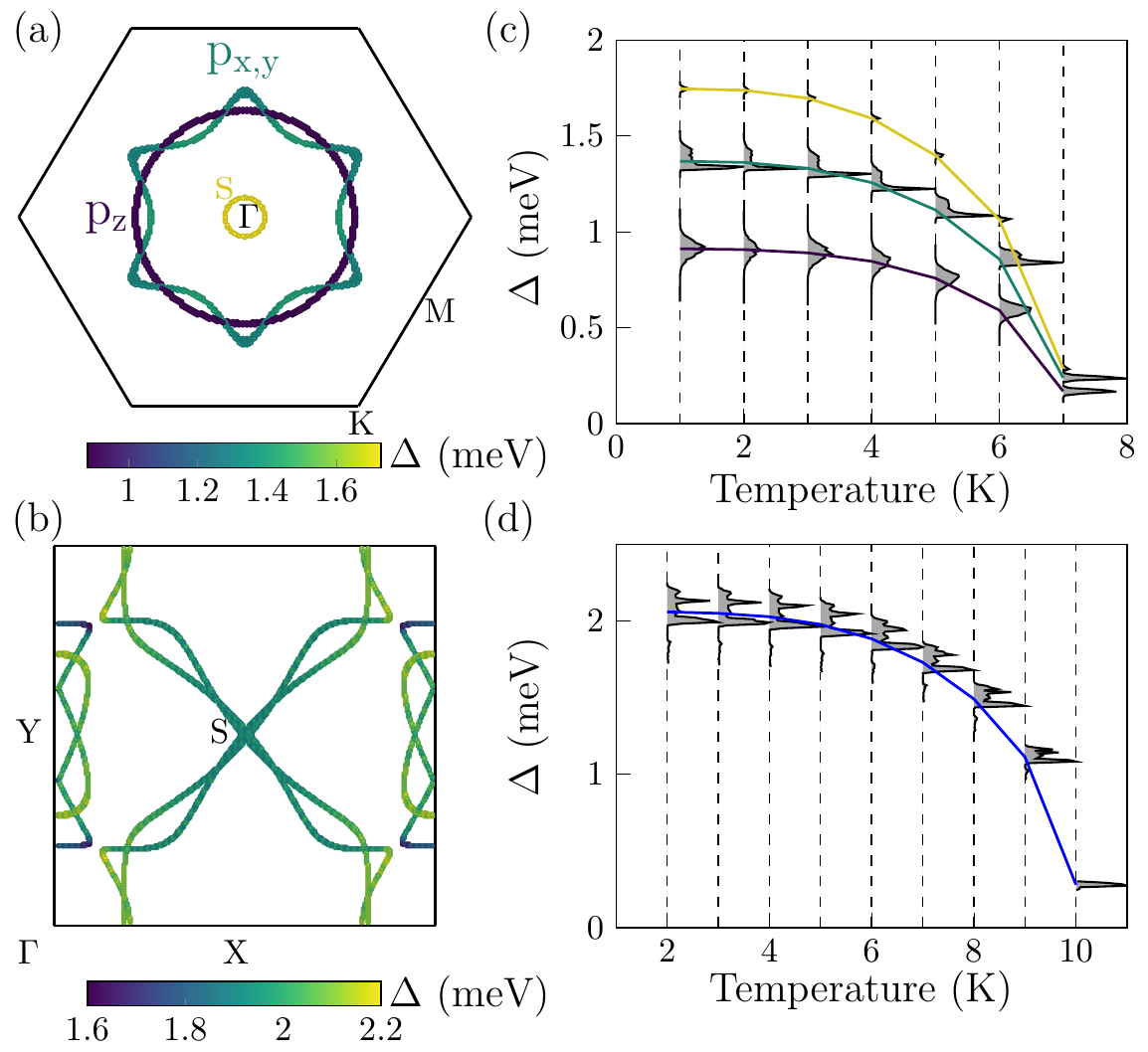}
\caption{Superconducting properties of gallenene. Momentum-dependent superconducting gap on the Fermi surface for (a) Ga-100 at 1 K and (b) Ga-010 at 4 K. Evolution of the superconducting gap distribution (gray areas) with temperature for (c) Ga-100 and (d) Ga-010. The weighted averages of the different domes are traced with solid lines.}
\label{fig3} 
\end{figure}

The EPC in both gallenene structures is decidedly strong enough to instigate superconductivity. To characterize their superconducting properties, we performed fully anisotropic Eliashberg calculations, as implemented in EPW \cite{PhysRevB.87.024505}. These calculations yield the superconducting gap as a function of Fermi wave vectors, $\Delta(\textbf{k}_F)$, indicating how particular electronic states of the Fermi surface contribute to the overall superconductivity in the monolayers. The results, shown in the Fig.~\ref{fig3}(a,b), reveal very different gap structures in the two gallenenes. As mentioned above and indicated in Fig.~\ref{fig3}(a), the three different sheets making up the Fermi surface of Ga-100 originate from pure $s$, $p_{x,y}$ and $p_{z}$ orbitals respectively. Even though the $p_{x,y}$ and $p_{z}$ sheets intersect in twelve points of the Brillouin zone, our analysis of the band character has shown that they do not hybridize, likely due to the planar structure of Ga-100. Each of these different sheets develops a distinct superconducting gap, making Ga-100 a rare three-gap superconductor -- a characteristic it shares with monolayer MgB$_2$ \cite{PhysRevB.96.094510}. The largest $\Delta$ values correspond to the Ga-$s$ states, amounting to 1.7 meV at the lowest temperatures, as shown in Fig.~\ref{fig2}(c). The intermediate gap values, of $\sim 1.4$ meV at low $T$, belong to the $p_{x,y}$ sheet, and the weakest gap of $\sim 0.8$ meV at low $T$ opens on the $p_{z}$ sheet. By solving the Eliashberg equations for a range of temperatures, as shown in Fig.~\ref{fig3}(c), we directly obtain that the gaps vanish at $\sim 7.5$ K, yielding the $T_c$ of Ga-100. The behavior of the superconducting gap of Ga-010 is very different, as shown in Fig.~\ref{fig3}(b). Here, no separate gaps can be distinguished on the Fermi surface, but rather a single anisotropic gap. This gap reaches 2.1 meV at low $T$, hence it is larger than the strongest gap of Ga-100, which results in a higher $T_c$ of 10 K for Ga-010 [as shown in Fig.~\ref{fig3}(d)].

\begin{figure}[t] 
\centering 
\includegraphics[width=\linewidth]{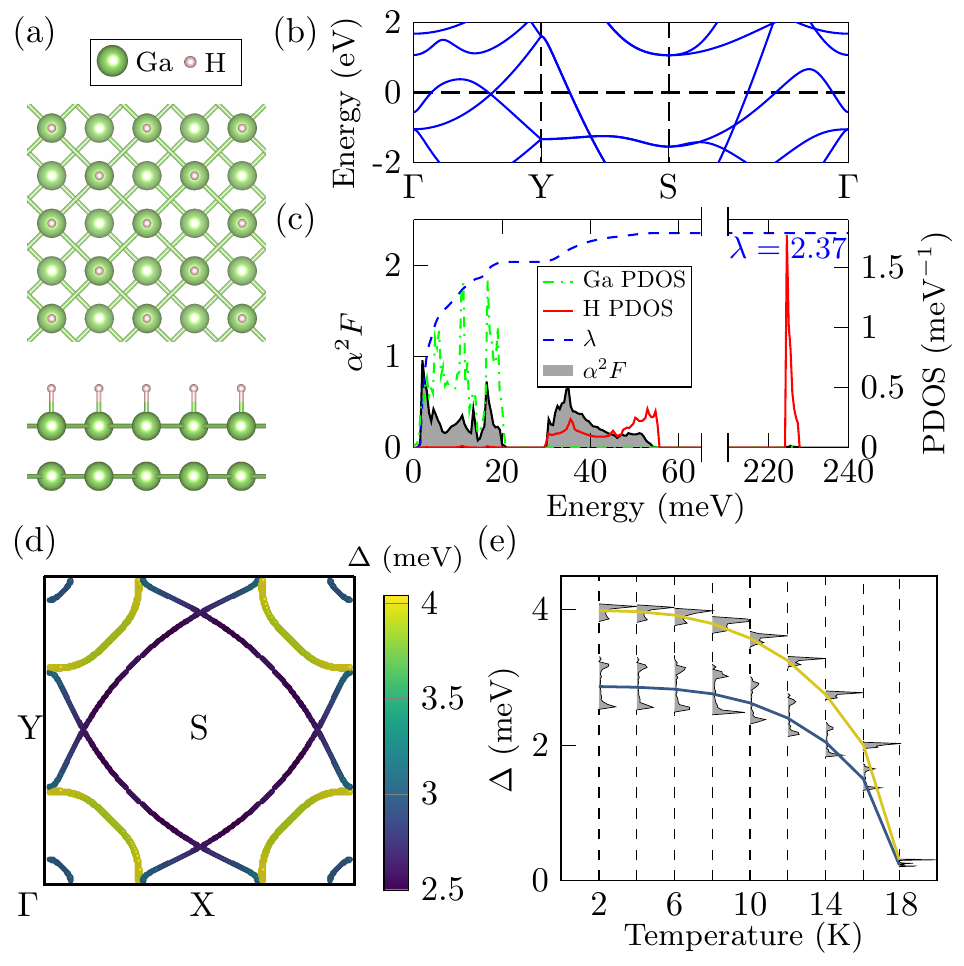}
\caption{Hydrogenated Ga-010. (a) Crystal structure: top and side view. (b) Electronic structure. (c) Isotropic Eliashberg function $\alpha^2F$, EPC $\lambda$ and atom-resolved phonon density of states (PHDOS). (d) Momentum-dependent superconducting gap on the Fermi surface. (e) Evolution of the superconducting gap distribution with temperature, including weighted averages of the different domes.} 
\label{fig4} 
\end{figure}

\paragraph{Effects of hydrogenation} As demonstrated for monolayer MgB$_2$ \cite{PhysRevB.96.094510,PhysRevLett.123.077001} and graphane \cite{PhysRevLett.105.037002}, addition of hydrogen (H) to a 2D material can be highly beneficial for its superconducting properties, owing to H's intrinsically high-energy vibrational modes and possible emergence of van Hove singularities in the electronic DOS, which both enhance the EPC \cite{PhysRevLett.123.077001}. Two hydrogenated forms of Ga-010, called gallenane, have been proposed \cite{Badalov-2018}. One of those compounds, featuring single-sided hydrogenation, stands out as particularly promising for superconductivity because of its high electronic DOS at $E_F$ (1.22 states/eV per atom). Furthermore, this gallenane structure is dynamically stable without any mechanical strain applied \cite{Badalov-2018}, unlike the freestanding gallenene structures. The crystal structure of monolayer gallenane also changes with respect to pure gallenene -- it is a square lattice comprising two layers of Ga atoms with H directly on top of every second column of Ga (yielding two H atoms in the unit cell), as shown in Fig.~\ref{fig4}(a).

The atom-resolved phonon DOS of gallenane [Fig.~\ref{fig4}(c)] shows that the H atoms strongly enhance the range of phononic energies, contributing in-plane modes of $25-55$ meV and out-of-plane modes of $220-230$ meV. Akin to pure gallenene, the Eliashberg function [Fig.~\ref{fig4}(c)] exhibits several peaks at low energies (up to 20 meV) due to Ga phonon modes, but also an additional dome stemming from the in-plane H modes. The latter contributes $19\%$ of the total EPC of 2.37 (compared with 2.07 for pure Ga-010). Hence, hydrogen plays a significant role in enhancing the EPC of this compound. Subsequently, we performed anisotropic Eliashberg calculations, which show the presence of two distinct gaps in gallenane [Fig.~\ref{fig4}(d)], making it a two-gap superconductor (compared to the single-gap nature of superconductivity in pure Ga-010). These two gaps correspond to electronic states with different spatial distribution, namely the stronger gap stems from a state that is localized in both Ga layers separately, while the electronic states harboring the weaker gap are mainly situated in the interlayer region (as shown in the Supplemental Material). Owing to the strong EPC in gallenane, we find the enhanced $T_c$ of 18 K, as shown in Fig.~\ref{fig4}(e). 

\paragraph{Conclusions} In summary, we have found two monolayer gallium (gallenene) structures to be superconducting, with $T_c$ in the range of $7-10$ K. One of them (Ga-100) has a planar honeycomb structure, hence represents the most elementary 2D superconductor known to date. This structural simplicity makes Ga-100 a versatile building block for various heterostructures, where interlayer hybridization with other elemental monolayers is likely to produce novel quantum phases. For example, gallenene--graphene heterostructures can be created owing to recent advances in growth of gallenene by intercalation of Ga between a graphene sheet and a substrate \cite{PhysRevMaterials.5.024006}. This system is expected to host a curious interplay between superconductivity and massless Dirac fermions. Combining gallenene with any of the heavy-element monolayers could enhance Ga-100's intrinsic SOC-induced topological states, demonstrated in this work. Furthermore, Cooper-pair tunneling across gallenene bilayers, or by extension in vertical heterostructures including a 2D barrier material (e.g., gallenene--graphene--gallenene), enables fabrication of ultrathin Josephson junctions \cite{Lee2015,Kim2017,Yabuki2016}, where multigap nature of superconductivity in Ga-100 enables exploration of additional functionalities through multiple Cooper-pair tunneling channels. Such junctions are envisaged as a key building block for applications in the field of quantum information and computing, and are likely to overcome the issues limiting the established bulk superconducting qubits, such as decoherence due to impurities at the interfaces \cite{Liu2019}. Finally, we have demonstrated that gallenene is prone to chemical functionalization, with a profound effect on its structural and electronic properties, and with the ability to tailor and to enhance its superconducting characteristics, as we exemplified on hydrogenated Ga-010. With such a unique blend of advantageous structural, superconducting and topological properties, and being tunable through chemical functionalization and hybridization in heterostructures, gallenene holds promise for major impact in both fundamental science and technology.

\begin{acknowledgments}
This work was supported by Research Foundation-Flanders (FWO) and the EU-COST action NANOCOHYBRI (Grant No. CA 16218). M.P. holds a FWO doctoral scholarship, and J.B. is a postdoctoral fellow of the FWO. The computational resources and services were provided by the VSC (Flemish Supercomputer Center), funded by the FWO and the Flemish Government -- department EWI.
\end{acknowledgments}

\bibliography{biblio}

\end{document}